\documentclass[a4paper]{jpconf}
\usepackage{graphicx}

\usepackage[centertags]{amsmath}
\usepackage{enumerate}
\usepackage{stmaryrd}
\usepackage{dsfont}
\usepackage{url}

\usepackage{amsfonts,amssymb,amsmath,amscd,amsthm}
\usepackage{mathrsfs}

\newtheorem{Definition}{Definition}
\newtheorem{Conjecture}{Conjecture}
\newtheorem{Example}{Example}
\newtheorem{Theorem}{Theorem}
\newtheorem{Claim}{Claim}
\newtheorem{Lemma}{Lemma}
\newtheorem{Remark}{Remark}
\newtheorem{Corollary}{Corollary}
\newtheorem{Proposition}{Proposition}

\newenvironment{Proof}{\textbf{Proof}}{ \qed \\}
\newenvironment{Thm}[1][]{\begin{Theorem} \normalfont \textbf{#1} \itshape}{\end{Theorem}}

\newenvironment{Def}[1][]{\begin{Definition} \normalfont \textbf{#1} \itshape}{\end{Definition}}
\newenvironment{Ex}[1][]{\begin{Example} \normalfont \textbf{#1}}{\qed \end{Example}}

\newenvironment{Rem}[1][]{\begin{Remark} \normalfont \textbf{#1}}{\end{Remark}}

\newcommand{\sN}{{\mathbb N}}

\newcommand{\sR}{{\mathbb R}}

\newcommand{\B}{\mathcal{B}}

\newcommand{\M}{\mathcal{M}}
\newcommand{\Pf}{\mathscr{P}}

\newcommand{\T}{\mathcal{T}}

\newcommand{\X}{\mathcal{X}}
\newcommand{\Y}{\mathcal{Y}}

\newcommand{\itDelta}{\mathit{\Delta}}
\newcommand{\itPi}{\mathit{\Pi}}

\begin{document}
\title{Time functions and $K$-causality between measures}

\author{Tomasz Miller${}^{1,2}$}

\address{${}^1$ Faculty of Mathematics and Information Science, Warsaw University of Technology, ul.~Koszykowa 75, 00-662 Warsaw, Poland}
\address{${}^2$ Copernicus Center for Interdisciplinary Studies, ul. Szczepa\'nska 1/5, 31-011 Krak\'ow, Poland}

\ead{T.Miller@mini.pw.edu.pl}

\begin{abstract}
Employing the notion of a coupling between measures, drawn from the optimal transport theory, we study the extension of the Sorkin--Woolgar causal relation $K^+$ onto the space $\Pf(\M)$ of Borel probability measures on a given spacetime $\M$. We show that Minguzzi's characterization of $K^+$ in terms of time functions possesses a ``measure-theoretic'' generalization. Moreover, we prove that the relation $K^+$ extended onto $\Pf(\M)$ retains its property of antisymmetry for $\M$ stably causal.
\end{abstract}

\section{Introduction}
In causality theory, i.e. the subfield of mathematical relativity studying the causal properties of spacetimes, the fundamental role is played by the binary relations of chronological and causal precedence called $I^+$ and $J^+$, respectively. An event $p$ is said to chronologically (causally) precede another event $q$ if they can be connected by means of a future-directed timelike (causal) curve, what is usually denoted $p \ll q$ ($p \preceq q$). For a concise review of causality theory we refer the reader to \cite{MS08}, whereas a more detailed exposition can be found e.g. in \cite{Beem,Penrose1972,BN83,HELargeScaleStructure,Wald}.

There are, however, other binary relations studied in causality theory, which offer an alternative way of modelling causal influence. One of the important examples is the relation $K^+$, introduced by Sorkin and Woolgar \cite{SorkinWoolgar} within the programme of reformulating causality theory along topological and order-theoretic lines, which in particular would encompass spacetimes with metrics of low regularity. $K^+$ was defined there as the smallest transitive and topologically closed relation containing $I^+$. Of course, if the spacetime $\M$ is causally simple, then $J^+$ is a closed subset of $\M^2$ and hence $K^+ = J^+$, but without the assumption of causal simplicity only the inclusion $K^+ \supseteq J^+$ holds, thus allowing for causal influence to involve more pairs of events than the causal curves alone can connect.

The Sorkin--Woolgar relation turns out to give some non-trivial insight into the property of \emph{stable causality}, as attested by the following result due to Minguzzi \cite{Minguzzi2009} (see also the article by Bernard and Suhr in the present volume):
\begin{Thm}
\label{MinguzziAS}
Spacetime $\M$ is stably causal iff it is $K$-causal, that is iff the relation $K^+$ is antisymmetric (and hence a partial order).
\end{Thm}

In addition, it has been shown (also by Minguzzi \cite{MinguzziUtilities}) that for $\M$ stably causal the relation $K^+$ can be completely characterized in terms of \emph{time functions}, i.e. continuous real-valued maps on $\M$, which are strictly increasing along every future-directed causal curve. It is worth noting that such maps exist precisely on stably causal spacetimes \cite{Sanchez05}. Concretely, the characterization is as follows.
\begin{Thm}
\label{MinguzziThm}
Let $\M$ be a stably causal spacetime and let $p, q \in \M$. Then
\begin{align}
\label{MinguzziT}
(p, q) \in K^+ \quad \Leftrightarrow \quad \textnormal{For every time function } t \quad t(p) \leq t(q).
\end{align}
\end{Thm}

Inspired by the causal structure in noncommutative Lorentzian geometry (\cite{CQG2013}, see also \cite{EcksteinUniverse17} and references therein) we have recently proposed a way to extend the causal precedence relation $J^+$ onto the space $\Pf(\M)$ of all Borel probability measures on a given spacetime $\M$ \cite{EcksteinMiller2015}. The proposed extension employs the notion of a \emph{coupling} of a pair of measures, drawn from the optimal transport theory \cite{UsersGuide,Villani2008} adapted to the Lorentzian setting, and can be successfully used to model the causal time-evolution of spatially distributed quantities from classical physics, such as charge or energy densities \cite[Section 2]{Miller16}. Surprisingly enough, it turns out to provide also a suitable framework for the study of quantum wave packets \cite{2NEW2016}.

Let us emphasize, however, that the method of extending $J^+$ onto $\Pf(\M)$ can be in fact applied to any (Borel) relation $R \subseteq \M^2$ (see Definition \ref{causality_def} below). In particular, in \cite{MillerUniverse17} we have studied the extension of the Sorkin--Woolgar relation $K^+$, addressing the question whether it retains its defining properties of transitivity and closedness, as well as establishing a ``measure-theoretic'' analogue of characterization (\ref{MinguzziT}). Although the first question has been answered positively, the characterization has been proven only under a stronger assumption of $\M$ being causally continuous. What is more, the problem whether $K^+$ extended onto $\Pf(\M)$ is still antisymmetric for $\M$ stably causal has been left for future investigations. 

The aim of the current paper is to fill both the above-mentioned gaps. To this end, we begin in Section \ref{sec::prelim} by providing the prerequisite definitions and some general results. Section \ref{sec::results} constitutes the main part of the article, with Theorem \ref{TimeFunctionsNowe} answering the question concerning the characterization with time functions, and Theorem \ref{ThmAS} tackling with the problem of antisymmetry. The (somewhat technical) proofs involve introducing and studying the so-called \emph{multi-time orderings}, which generalize time orderings employed in \cite{MinguzziUtilities}. 

\section{Preliminaries}
\label{sec::prelim}
From now on, the term ``measure'' will always stand for ``Borel probability measure''.

Given a pair of measures $\mu,\nu \in \Pf(\M)$, we call $\omega \in \Pf(\M^2)$ a \emph{coupling} of $\mu$ and $\nu$ if the latter two measures are $\omega$'s marginals, that is if $\pi^1_\sharp \omega = \mu$ and $\pi^2_\sharp \omega = \nu$, where $\pi^i: \M^2 \rightarrow \M$, $i=1,2$ denote the canonical projections. We shall denote the set of all such couplings by $\itPi(\mu,\nu)$. Borrowing the terminology from Suhr \cite[Definition 2.4]{Suhr2016}, let us put forward the following general definition:
\begin{Def}
\label{causality_def}
Let $\M$ be a Polish space and let $R \subseteq \M^2$ be a \emph{Borel} binary relation. For any $\mu,\nu \in \Pf(\M)$ we say that $\mu$ is \emph{$R$-related} with $\nu$ if there exists $\omega \in \itPi(\mu,\nu)$ such that $\omega(R) = 1$.
\end{Def}

\begin{Ex}
\label{ExEq}
As a simple illustration of the above concept, consider the \emph{equality relation} $=$, which, when regarded as a subset of $\M^2$, is nothing but $\itDelta(\M)$, where $\itDelta: \M \rightarrow \M^2$, $\itDelta(p) = (p,p)$ denotes the diagonal map. Then for any $\mu, \nu \in \Pf(\M)$ the existence of $\omega \in \itPi(\mu,\nu)$ such that $\omega(\itDelta(\M)) = 1$ is a necessary and sufficient condition for $\mu$ and $\nu$ to be actually equal. 

Indeed, necessity follows from the observation that, for any $\mu \in \Pf(\M)$, $\itDelta_\sharp \mu \in \itPi(\mu,\mu)$ and of course $\itDelta_\sharp \mu(\itDelta(\M)) = \mu(\itDelta^{-1}(\itDelta(\M))) = \mu(\M) = 1$. For the proof of sufficiency, consult \cite[Lemma 4]{EcksteinMiller2015}.
\end{Ex}

By analogy with $J^+$, for any $\X \subseteq \M$ we introduce the notation $R^+(\X) := \pi^2\left((\X \times \M) \cap R\right)$ and $R^-(\X) := \pi^1\left((\M \times \X) \cap R\right)$. Notice that the sets $R^\pm(\X)$ need not be Borel even if $\X$ is a Borel set. Nevertheless, being projections of Borel sets, they are \emph{universally measurable}, which means that for any measure $\mu \in \Pf(\M)$ the sets $R^\pm(\X)$ are Borel up to a $\mu$-negligible set, and therefore the quantity $\mu(R^\pm(\X))$ is well defined \cite{Aliprantis}.

We will need the following powerful characterization of $R$-relatedness due to Suhr \cite[Theorem 2.5]{Suhr2016}.
\begin{Thm}
\label{Suhrowe}
Let $\M$ be a Polish space and let the relation $R \subseteq \M^2$ be closed (topologically). Then for any $\mu, \nu \in \Pf(\M)$ the following conditions are equivalent
\begin{enumerate}[(i)]
\item $\mu$ is $R$-related with $\nu$.
\item For any Borel $\B \subseteq \M$
\begin{align*}
\mu(\B) \leq \nu(R^+(\B)) \quad \textnormal{and} \quad \mu(R^-(\B)) \geq \nu(\B).
\end{align*}
\end{enumerate}
\end{Thm}

In this paper we will be dealing with \emph{closed preorders}, i.e. those relations $R \subseteq \M^2$ which are reflexive, transitive and topologically closed. We will write $\mu \preceq_R \nu$ to express the $R$-relatedness of $\mu$ and $\nu$. 
Notice that for any compact $C \subseteq \M$ the sets $R^\pm(C)$ are closed.

As a first general result, we obtain the following alternative characterizations of $\preceq_R$.
\begin{Thm}
\label{main}
Let $\M$ be a Polish space and let $R \subseteq \M^2$ be a closed preorder. For any $\mu,\nu \in \Pf(\M)$ the following conditions are equivalent:
\begin{enumerate}[1{$^\bullet$}]
\item $\mu \preceq_R \nu$.
\item For any compact subset $\ C \subseteq \M$
\begin{align}
\label{main1}
\mu(R^+(C)) \leq \nu(R^+(C)).
\end{align}
\item For any Borel subset $\X \subseteq \M$ such that $R^+(\X) \subseteq \X$
\begin{align}
\label{main2}
\mu(\X) \leq \nu(\X).
\end{align}
\end{enumerate}
Additionally, conditions $2^\bullet$ and $3^\bullet$ are equivalent with their ``past'' counterparts:
\begin{enumerate}[1{$^{\prime \bullet}$}]
\setcounter{enumi}{1}
\item For any compact subset $\ C \subseteq \M$
\begin{align}
\label{main1prime}
\mu(R^-(C)) \geq \nu(R^-(C)).
\end{align}
\item For any Borel subset $\Y \subseteq \M$ such that $R^-(\Y) \subseteq \Y$
\begin{align}
\label{main2prime}
\mu(\Y) \geq \nu(\Y).
\end{align}
\end{enumerate}
\end{Thm}
\begin{Proof}\textbf{.}
We adapt here of the first part of the proof of \cite[Theorem 2]{MillerUniverse17}.

$1^\bullet \ \Rightarrow \ 2^\bullet$ By the closedness of $R$, for any compact $C \subseteq \M$ the set $R^+(C)$ is closed and hence Borel. Denoting its characteristic function by $\chi$ (which is a Borel map), the inequality $\chi(p) \leq \chi(q)$ holds for all $(p,q) \in R$ by transitivity. Finally, on the strength of $1^\bullet$, there exists $\omega \in \itPi(\mu,\nu)$ supported on $R$. Altogether, one can write that
\begin{align*}
\mu(R^+(C)) & = \int\limits_\M \chi(p) d\mu(p) = \int\limits_{\M^2} \chi(p) d\omega(p,q) = \int\limits_{R} \chi(p) d\omega(p,q) \leq \int\limits_{R} \chi(q) d\omega(p,q)
\\
& = \int\limits_{\M^2} \chi(q) d\omega(p,q) = \int\limits_\M \chi(q) d\nu(q) = \nu(R^+(C)).
\end{align*}
One similarly proves that $1^\bullet \ \Rightarrow \ 2^{\prime \bullet}$.

$2^\bullet \ \Rightarrow \ 3^\bullet$ Let the set $\X$ be as specified in $3^\bullet$. Taking any compact $C \subseteq \X$, we have that $R^+(C) \subseteq R^+(\X) \subseteq \X$ and hence
\begin{align*}
\mu(C) \leq \mu(R^+(C)) \leq \nu(R^+(C)) \leq \nu(\X),
\end{align*}
where in the second inequality we have used $2^\bullet$. Using the fact that $\mu$, being a Borel measure on a Polish space, is inner regular (a.k.a. \emph{tight} \cite[Lemma 12.6]{Aliprantis}), we hence obtain that
\begin{align*}
\mu(\X) = \sup \{ \mu(C) \ | \ C \subseteq \X \textnormal{ compact} \} \leq \nu(\X).
\end{align*}
One similarly proves that $2^{\prime \bullet} \ \Rightarrow \ 3^{\prime \bullet}$.

$3^\bullet \ \Leftrightarrow \ 3^{\prime \bullet}$ Denote $\X^c := \M \setminus \X$. The possibility of moving between these two conditions relies on the obvious equality $\mu(\X^c) = 1 - \mu(\X)$ valid for any $\mu \in \Pf(\M)$ and any Borel set $\X \subseteq \M$, and on the following equivalence of inclusions: $R^+(\X) \subseteq \X \ \Leftrightarrow \ R^-(\X^c) \subseteq \X^c$. The latter has been stated in \cite[Proposition 1]{EcksteinMiller2015} for $J^+$, but the proof conducted there is in fact valid for \emph{any} relation $R$.

$3^\bullet, \, 3^{\prime \bullet} \ \Rightarrow \ 1^\bullet$ It suffices to notice that condition \emph{(ii)} in Theorem \ref{Suhrowe} is satisfied.

Indeed, the first inequality in \emph{(ii)} follows directly from $3^\bullet$ with $\X := R^+(\B)$ (up to a $\mu$-negligible set) and the obvious inequality $\mu(\B) \leq \mu(R^+(\B))$. Similarly, the second inequality follows from $3^{\prime \bullet}$ with $\Y := R^-(\B)$ (up to a $\nu$-negligible set) and another self-evident inequality $\nu(R^-(\B)) \geq \nu(\B)$.
\end{Proof}

\section{Main results}
\label{sec::results}

In \cite{MillerUniverse17}, we have provided several characterizations of the $K$-causality relation between measures, some of which have been proven to hold for all stably causal spacetimes, whilst others seemed to demand a stronger requirement of causal continuity. Below, however, we show that the latter requirement is in fact redundant. In other words, we upgrade \cite[Theorem 2]{MillerUniverse17} to the following result.

\begin{Thm}
\label{TimeFunctionsNowe}
Let $\M$ be a stably causal spacetime. For any $\mu,\nu \in \Pf(\M)$ the following conditions are equivalent.
\begin{enumerate}[1{$^\bullet$}]
\item $\mu \preceq_K \nu$.
\setcounter{enumi}{3}
\item For any time function $t$ and any $\lambda \in \sR$
\begin{align}
\label{main3}
\mu\left(t^{-1}((\lambda, +\infty))\right) \leq \nu\left(t^{-1}((\lambda, +\infty))\right).
\end{align}
\item For any bounded time function $t$
\begin{align}
\label{main4}
\int_\M t d\mu \leq \int_\M t d\nu.
\end{align}
\end{enumerate}
\end{Thm}

The numbers $2^\bullet$, $3^\bullet$ have been omitted here as they refer to the conditions listed in Theorem \ref{main}, which of course holds in the special case $R := K^+$.

Additionally, the following two remarks from \cite{MillerUniverse17} are still valid (with their proofs unchanged).
\begin{Rem}
\label{rem1}
Without loss of (or gain in) generality, in condition $5^\bullet$ the term ``bounded'' can be replaced with ``$\mu$- and $\nu$-integrable'', whereas the term ``time'' can be substituted with ``temporal'', ``smooth time'', ``smooth causal'' or ``continuous causal''.
\end{Rem}

\begin{Rem}
\label{rem2}
Condition $4^\bullet$ can equivalently employ the \emph{closed} half-lines. In other words, the following condition is equivalent to $4^\bullet$:
\begin{enumerate}[1{$^{\prime \bullet}$}]
\setcounter{enumi}{3}
\item For any time function $t$ and any $\lambda \in \sR$
\begin{align}
\label{main3a}
\mu\left(t^{-1}([\lambda, +\infty))\right) \leq \nu\left(t^{-1}([\lambda, +\infty))\right).
\end{align}
\end{enumerate}
\end{Rem}

In order to prepare the ground for the proof of Theorem \ref{TimeFunctionsNowe}, we begin by slightly strengthening Minguzzi's Theorem \ref{MinguzziThm}.
\begin{Thm}
\label{MinguzziThmMod}
Let $\M$ be a stably causal spacetime. Then there exists a \emph{countable} family of \emph{bounded} time functions $\{t_\alpha: \M \rightarrow (0,1) \}_{\alpha \in \sN}$, such that for any $p,q \in \M$
\begin{align}
\label{MinguzziTMod}
(p, q) \in K^+ \ \quad \Leftrightarrow \quad \ \forall \, \alpha \in \sN \quad t_\alpha(p) \leq t_\alpha(q)
\end{align}
and, moreover,
\begin{align}
\label{CorMinguzziTMod}
    \left[ \forall \, \alpha \in \sN \quad t_\alpha(p) = t_\alpha(q) \right] \quad \Leftrightarrow \quad p = q.
\end{align}
\end{Thm}
\begin{Proof}\textbf{.}
The implication `$\Rightarrow$' in (\ref{MinguzziTMod}) follows trivially from Theorem \ref{MinguzziThm}. In order to show the converse, we will demonstrate how to construct a countable family of time functions $\{t_\alpha: \M \rightarrow (0,1) \}_{\alpha \in \sN}$ such that, for any $p,q \in \M$
\begin{align*}
    (p,q) \notin K^+ \quad \Rightarrow \quad \exists \, \alpha \in \sN \quad t_\alpha(p) > t_\alpha(q).
\end{align*}

To begin with, consider the open set $\M^2 \setminus K^+$. By Theorem \ref{MinguzziThm}, for any pair $(p,q) \in \M^2 \setminus K^+$ we can pick a time function $t_{p,q}$ such that $t_{p,q}(p) - t_{p,q}(q) > 0$. Notice now that the map $(t_{p,q} \circ \pi^1 - t_{p,q} \circ \pi^2): \M^2 \rightarrow \sR$ is continuous, and therefore the following family of inverse images:
\begin{align*}
    \left\{ (t_{p,q} \circ \pi^1 - t_{p,q} \circ \pi^2)^{-1}((0,+\infty)) \right\}_{(p,q) \in M^2 \setminus K^+}
\end{align*}
constitutes an open cover of $\M^2 \setminus K^+$. Since the latter is a separable metric space (being an open subspace of a separable metric space $\M^2$), it possesses the Lindel\"{o}f property \cite[Theorem 16.11]{Willard}, i.e. we can choose a countable subcover of the above cover:
\begin{align*}
    \left\{ (t_{p_\alpha,q_\alpha} \circ \pi^1 - t_{p_\alpha,q_\alpha} \circ \pi^2)^{-1}((0,+\infty)) \right\}_{\alpha \in \sN}.
\end{align*}
To obtain the desired countable family of bounded time functions, it now suffices to define $t_\alpha := \varphi \circ t_{p_\alpha,q_\alpha}$ for every $\alpha \in \sN$, where $\varphi \in C_b(\sR)$ is a fixed continuous strictly increasing function attaining values in $(0,1)$ (for instance, $\varphi(x) := \tfrac{1}{2} + \tfrac{1}{2} \tanh x$).

As for condition (\ref{CorMinguzziTMod}), it follows directly from (\ref{MinguzziTMod}) and Theorem \ref{MinguzziAS}.
\end{Proof}

In \cite{MillerUniverse17} it was proven that the Sorkin--Woolgar relation retains its defining properties of transitivity and closedness when extended onto measures. With the help of Theorem \ref{MinguzziThmMod}, we can show that this extension is also antisymmetric (and hence a partial order) provided $\M$ is stably causal. In other words, Minguzzi's Theorem \ref{MinguzziAS} still holds in the more general setting of $K$-causality between measures.
\begin{Thm}
\label{ThmAS}
Let $\M$ be a~stably causal spacetime. Then the relation $\preceq_K$ on $\Pf(\M)$ is antisymmetric.
\end{Thm}
\begin{Proof}\textbf{.}
The major part of the proof can be straightforwardly adapted from that of \cite[Theorem 12]{EcksteinMiller2015}, where it was proven that the relation $J^+$ (extended onto $\Pf(\M)$) is antisymmetric under some mild assumptions on the causal properties of $\M$. In fact, there is only one step of that proof which requires a nontrivial modification. Namely, we need to show here that, for any fixed $\mu \in \Pf(\M)$, the only $\omega \in \itPi(\mu,\mu)$ satisfying $\omega(K^+) = 1$ is $\itDelta_\sharp \mu$. On the strength of \cite[Lemma 4]{EcksteinMiller2015}, it suffices to prove that $\omega(\itDelta(\M)) = 1$.

To this end, observe first that for any $f \in C_b(\M)$ we have
\begin{align}
\label{ThmAS1}
    \int_{K^+} (f(q) - f(p)) d\omega(p,q) = 0,
\end{align}
what can be obtained by subtracting the identities $\int_\M f(p) d\mu(p) = \int_{K^+} f(p) d\omega(p,q)$ and $\int_\M f(q) d\mu(q) = \int_{K^+} f(q) d\omega(p,q)$, true by the very assumption on $\omega$.

Let us now fix a countable family of time functions $\{t_\alpha: \M \rightarrow (0,1) \}_{\alpha \in \sN}$ provided by Theorem \ref{MinguzziThmMod}. We claim that the following chain of equalities is true
\begin{align}
\nonumber
    1 = \omega(K^+) & = \omega\left( \left\{ (p,q) \in K^+ \ \Big| \ \sum_{\alpha = 1}^{\infty} 2^{-\alpha} \, t_\alpha(p) = \sum_{\alpha = 1}^{\infty} 2^{-\alpha} \, t_\alpha(q) \right\} \right)
\\
\label{ThmAS2}
 & = \omega\left( \bigcap_{\alpha = 1}^{\infty} \left\{ (p,q) \in K^+ \ | \ t_\alpha(p) = t_\alpha(q) \right\} \right) = \omega(\itDelta(\M)).
\end{align}

Indeed, the first equality follows from the very definition of $\omega$. To see why the second equality holds, observe that the complementary subset $\{ (p,q) \in K^+ \ | \ \sum_{\alpha = 1}^{\infty} 2^{-\alpha} \, t_\alpha(p) < \sum_{\alpha = 1}^{\infty} 2^{-\alpha} \, t_\alpha(q) \}$ is $\omega$-null on the strength of (\ref{ThmAS1}), in which we have plugged $f := \sum_{\alpha = 1}^{\infty} 2^{-\alpha} \, t_\alpha$. As for the third equality, one can actually prove the equality of the \emph{sets} involved. In order to show the nontrivial inclusion ``$\subseteq$'', take any $(p,q) \in K^+$ such that $\exists \, \alpha_0 \in \sN \ t_{\alpha_0}(p) < t_{\alpha_0}(q)$. Of course, since $t_\alpha$'s are all time functions, they satisfy $t_\alpha(p) \leq t_\alpha(q)$ by Theorem \ref{MinguzziT}, and therefore we obtain that $\sum_{\alpha = 1}^{\infty} 2^{-\alpha} \, t_\alpha(p) < \sum_{\alpha = 1}^{\infty} 2^{-\alpha} \, t_\alpha(q)$. Finally, the fourth equality follows from condition (\ref{CorMinguzziTMod}).
\end{Proof}

In the proof of Theorem \ref{TimeFunctionsNowe} we will need another kind of causal relation, which generalizes Minguzzi's time orderings \cite{MinguzziUtilities}. Recall that, given a time function $t: \M \rightarrow \sR$, its corresponding \emph{time ordering} is a (closed total) preorder defined as
\begin{align*}
T^+[t] = \left\{ (p,q) \in \M^2 \, | \ t(p) \leq t(q) \right\}.
\end{align*}
We generalize this definition onto finite collections of time functions in a straightforward way.
\begin{Definition}
Let $\M$ be a stably causal spacetime and let $F:= \{t_1,t_2,\ldots,t_n\}$, $n \in \sN$ be a finite set of time functions. Its corresponding \emph{multi-time ordering} is a (closed) preorder defined via
\begin{align*}
T^+[F] := \bigcap\limits_{t \in F} T^+[t] = \left\{ (p,q) \in \M^2 \, | \ t_\alpha(p) \leq t_\alpha(q), \ \alpha = 1,\ldots,n \right\}
\end{align*}
\end{Definition}

\begin{Thm}
\label{multiThm}
Let $\M, F$ and $n$ be as above. Then for any $\mu, \nu \in \Pf(\M)$ the following conditions are equivalent:
\begin{enumerate}[1{$^\bullet$}]
\item $\mu \preceq_{T[F]} \nu$.
\item For any $\Phi \in C_b(\sR^n)$ componentwise increasing
\begin{align}
\label{mainF}
\int_\M \Phi(t_1,\ldots,t_n) d\mu \leq \int_\M \Phi(t_1,\ldots,t_n) d\nu.
\end{align}
\end{enumerate}
\end{Thm}
\begin{Proof}\textbf{.}
Implication $1^\bullet \Rightarrow 2^\bullet$ can be proven through the following chain of (in)equalities, valid for any componentwise increasing function $\Phi \in C_b(\sR^n)$:
\begin{align*}
    \int_\M \Phi(t_1,\ldots,t_n) d\mu & = \int_{K^+} \Phi(t_1(p),\ldots,t_n(p)) d\omega(p,q)
    \\
    & \leq \int_{K^+} \Phi(t_1(q),\ldots,t_n(q)) d\omega(p,q) = \int_\M \Phi(t_1,\ldots,t_n) d\nu,
\end{align*}
where the equalities hold by the very definition of $\preceq_K$, whereas the inequality is true by Theorem~\ref{MinguzziThm} and the assumption on $\Phi$.

In order to show the converse implication, we will demonstrate that $2^\bullet$ implies the following condition
\begin{align}
    \label{mainFcel}
    \mu(T^+[F](C)) \leq \nu(T^+[F](C)) \quad \textrm{for any compact subset } C \subseteq \M,
\end{align}
which is equivalent to $1^\bullet$ by Theorem \ref{main}.

Let us first assume that $C$ is \emph{finite}. In order to show that inequality (\ref{mainFcel}) holds for $C = \{q_1, \ldots, q_s\}$, consider the maps $\T_{k,l}: \M \rightarrow \sR$ defined, for any $k,l \in \sN$, via
\begin{align*}
    \T_{k,l}(p) := \varphi_l^- \left( \sum_{i=1}^s \prod_{\alpha=1}^n \varphi_k^+\left(t_\alpha(p) - t_\alpha(q_i)\right) \right),
\end{align*}
where $\varphi^\pm_k \in C_b(\sR)$ are defined as $\varphi^\pm_k(x) := \tfrac{1}{2} + \tfrac{1}{2} \tanh(k^2 x \pm k)$. Observe that the sequence $(\varphi^+_k)$ is pointwise convergent to the characteristic function of the \emph{closed} half-line $\chi_{[0,+\infty)}$, whereas the pointwise limit of the sequence $(\varphi^-_l)$ is the characteristic function of the \emph{open} half-line $\chi_{(0,+\infty)}$. But this actually means that, for any $p \in \M$,
\begin{align*}
    \lim\limits_{l \rightarrow +\infty} \lim\limits_{k \rightarrow +\infty} \T_{k,l}(p) = \left\{ \begin{array}{ll}
        1 & \textrm{if } \ \exists \, i \in \{1,\ldots,s\} \ \forall \, \alpha \in \{1,\ldots,n\} \quad t_\alpha(p) \geq t_\alpha(q_i) \\
        0 & \textrm{otherwise.}
    \end{array}
    \right.
\end{align*}
In other words, the iterated pointwise limit of the sequence $(\T_{k,l})$ calculated first with respect to $k$ and then $l$ is nothing but the characteristic function of the set $T^+[F](C)$.

Observe now that for any $k,l \in \sN$ the map $\T_{k,l}$ is of the form $\Phi(t_1,\ldots,t_n)$ with $\Phi \in C_b(\sR^n)$ componentwise increasing. On the strength of $2^\bullet$, we thus have that
\begin{align*}
\int_\M \T_{k,l} \, d\mu \leq \int_\M \T_{k,l} \, d\nu.
\end{align*}
Invoking Lebesgue's dominated convergence theorem twice, we pass first to the limit $k \rightarrow +\infty$ and then to the limit $l \rightarrow +\infty$, thus obtaining (\ref{mainFcel}) for $C$ finite.

Suppose now that $C \subseteq \M$ is \emph{any} compact subset. Our aim now is to construct a sequence $(C_m)_{m \in \sN}$ of finite subsets of $\M$ such that
\begin{align*}
    \forall \, m \in \sN \quad T^+[F](C_{m+1}) \subseteq T^+[F](C_{m}) \quad \textrm{and} \quad \bigcap\limits_{m = 1}^\infty T^+[F](C_{m}) = T^+[F](C),
\end{align*}
as this would already complete the proof of (\ref{mainFcel}) via
\begin{align*}
    \mu(T^+[F](C)) & = \mu\left( \bigcap\limits_{m = 1}^\infty T^+[F](C_m) \right) = \lim\limits_{m \rightarrow +\infty} \mu\left(T^+[F](C_{m})\right) 
    \\
    & \leq \lim\limits_{m \rightarrow +\infty} \nu\left(T^+[F](C_{m})\right) = \nu\left( \bigcap\limits_{m = 1}^\infty T^+[F](C_m) \right) = \nu(T^+[F](C)).
\end{align*}
We shall construct the sequence $(C_m)$ recursively. To this end, let us first introduce the following open subsets: $B(C,R) := \bigcup_{x \in C} B(x,R)$ (for any $R>0$) and, for any $p \in \M$,
\begin{align*}
    V(p) := \bigcap\limits_{\alpha = 1}^n t_\alpha^{-1}\left( (t_\alpha(p), +\infty) \right) = \left\{ q \in \M \ | \ \forall \, \alpha \in \{1,\ldots,n\} \ \ t_\alpha(p) < t_\alpha(q) \right\}.
\end{align*}
Two immediate observations follow. First, for any $R>0$ the family $\{V(p)\}_{p \in B(C,R)}$ constitutes an open cover of $C$ (and hence it possesses a finite subcover). Indeed, for any $q \in C$ one can always choose $p \in B(q,R) \setminus \{q\}$ which causally precedes $q$ and thus $t(p) < t(q)$ for any time function $t$. Second, it is easy to notice that $\overline{V(p)} = T^+[F](p)$.

To begin the construction of $(C_m)$, let $R_1 := 1$ and let the family $\{V(p^{(1)}_1), \ldots, V(p^{(1)}_{s_1})\}$ be a finite subcover of the open cover $\{V(p)\}_{p \in B(C,R_1)}$ of $C$. Define $C_1 := \{p^{(1)}_1, p^{(1)}_2, \ldots, p^{(1)}_{s_1}\}$.

Suppose now we have constructed $C_m = \{p^{(m)}_1, \ldots, p^{(m)}_{s_m}\}$ for some $m \in \sN$. In order to construct $C_{m+1}$, define first
\begin{align*}
    R_{m+1} := \tfrac{1}{2}\min \left\{ \textrm{dist}\left(C, \ \M \setminus \bigcup_{i=1}^{s_m} V(p_i^{(m)})\right), R_m \right\},
\end{align*}
where the distance (calculated with respect to some fixed auxiliary Riemannian metric on $\M$) is positive since $C$ is compact and $\M \setminus \bigcup_{i=1}^{s_m} V(p_i^{(m)})$ is closed. Similarly as in the initial step, consider now a finite subcover $\{V(p^{(m+1)}_1), \ldots, V(p^{(m+1)}_{s_{m+1}})\}$ of the open cover $\{V(p)\}_{p \in B(C,R_{m+1})}$ of $C$ and define $C_{m+1} := \{p^{(m+1)}_1, \ldots, p^{(m+1)}_{s_{m+1}}\}$.

Let us show that $T^+[F](C_{m+1}) \subseteq T^+[F](C_{m})$ for any $m \in \sN$. Indeed, take any $q \in T^+[F](C_{m+1})$, i.e. assume there exists $j \in \{1,\ldots,s_{m+1}\}$ such that $t_\alpha(p_j^{(m+1)}) \leq t_\alpha(q)$ for $\alpha = 1,\ldots,n$. Notice that, by construction,
\begin{align*}
p_j^{(m+1)} \in B(C,R_{m+1}) \, \subseteq \, \bigcup_{i=1}^{s_m} V(p_i^{(m)}) \, \subseteq \, \bigcup_{i=1}^{s_m} \overline{V(p_i^{(m)})} = T^+[F](C_{m}),
\end{align*}
and therefore, altogether, there exists $i \in \{1,\ldots,s_m\}$ such that for any $\alpha \in \{1,\ldots,n\}$ $t_\alpha(p_i^{(m)}) \leq t_\alpha(p_j^{(m+1)}) \leq t_\alpha(q)$. Hence $q \in T^+[F](C_{m})$.

Finally, it remains to prove that $\bigcap_{m = 1}^\infty T^+[F](C_{m}) = T^+[F](C)$. In order to prove ``$\subseteq$'', take any $q \in \M$ satisfying
\begin{align*}
    \forall \, m \in \sN \ \ \exists \, p_m \in C_m \ \ \forall \, \alpha \in \{1,\ldots,n\} \quad t_\alpha(p_m) \leq t_\alpha(q).
\end{align*}
By construction, for any $m \in \sN$ one has $\textrm{dist}(C,p_m) < R_m \leq 2^{m-1} \leq 1$. This implies that the sequence $(p_m)$, being contained in a precompact set $B(C,1)$, has a subsequence convergent to some $p$, which actually must lie in $\bigcap_{m = 1}^\infty B(C,2^{m-1}) = \overline{C} = C$. By the continuity of time functions, this means that $t_\alpha(p) \leq t_\alpha(q)$ for $\alpha=1,\ldots,n$ and so $q \in T^+[F](C)$.

To obtain the converse inclusion ``$\supseteq$'', fix any $m \in \sN$ and recall that the family $\{V(p_i^{(m)})\}_{i=1}^{s_m}$ covers $C$, hence
\begin{align*}
    C \, \subseteq \, \bigcup_{i=1}^{s_m} V(p_i^{(m)}) \, \subseteq \, \bigcup_{i=1}^{s_m} \overline{V(p_i^{(m)})} \, = \, T^+[F](C_{m}).
\end{align*}
By transitivity, this implies that $T^+[F](C) \subseteq T^+[F](T^+[F](C_m)) = T^+[F](C_m)$.
\end{Proof}

With the above characterization of multi-time orderings between measures at hand, we are finally ready to prove the characterization of $\preceq_K$ by means of time functions for $\M$ stably causal.
\\

\begin{Proof}\textbf{ of Theorem \ref{TimeFunctionsNowe}.} Implications $1^\bullet \Rightarrow 4^\bullet \Rightarrow 5^\bullet$ have been established in \cite{MillerUniverse17}, whereas the remaining implication $5^\bullet \Rightarrow 1^\bullet$ has been proven there only under the additional assumption of the causal continuity of $\M$. In order to show that this implication holds in all stably causal spacetimes, let us first fix a countable family of time functions $\{t_\alpha\}_{\alpha \in \sN}$ with the property (\ref{MinguzziTMod}). On the strength of $5^\bullet$ and Theorem \ref{multiThm} we obtain that for any $n \in \sN$ there exists $\omega_n \in \itPi(\mu,\nu)$ such that
\begin{align*}
    \omega_n\left(T^+[\{t_1,t_2,\ldots,t_n\}]\right) = \omega_n\left( \bigcap_{\alpha = 1}^n T^+[t_\alpha] \right) = 1.
\end{align*}
By the narrow compactness of $\itPi(\mu,\nu)$ (cf. \cite[proof of Theorem 1.5]{UsersGuide}), the sequence $(\omega_n)_{n \in \sN}$ has a subsequence $(\omega_{n_l})_{l \in \sN}$ narrowly convergent to some $\omega \in \itPi(\mu,\nu)$. We claim that $\omega(K^+) = 1$, what would already yield $1^\bullet$.

Indeed, observe first that for any $k,n \in \sN$
\begin{align*}
    k \leq n \quad \Rightarrow \quad \omega_n\left( \bigcap_{\alpha = 1}^k T^+[t_\alpha] \right) \geq \omega_n\left( \bigcap_{\alpha = 1}^n T^+[t_\alpha] \right) = 1.
\end{align*}
Therefore, by the portmanteau theorem \cite[Chapter 18, Theorem 6]{ModProb},
\begin{align*}
    \forall \, k \in \sN \quad 1 \geq \omega\left( \bigcap_{\alpha = 1}^k T^+[t_\alpha]\right) \geq \limsup\limits_{l \rightarrow +\infty} \, \omega_{n_l}\left( \bigcap_{\alpha = 1}^k T^+[t_\alpha] \right) \geq 1,
\end{align*}
and hence
\begin{align*}
    \omega(K^+) = \omega\left( \bigcap_{\alpha = 1}^\infty T^+[t_\alpha]\right) = \lim\limits_{k \rightarrow +\infty} \omega\left( \bigcap_{\alpha = 1}^k T^+[t_\alpha]\right) = 1,
\end{align*}
where in the first equality we have used property (\ref{MinguzziTMod}).
\end{Proof}

\ack
The author is grateful to Micha{\l} Eckstein for the careful perusal of the manuscript as well as to the Referees for their insightful comments. The research was made possible through the support of the National Science Centre, Poland via the grant Sonatina 2017/24/C/ST2/00322.

\section*{References}
\bibliography{causality}

\end{document}